\documentclass[twoside,twocolumn,tightenlines, showpacs,aps,prb]{revtex4}
\usepackage{graphicx}
\newcommand{\be}{\begin{equation}}
\newcommand{\ee}{\end{equation}}

\begin{document}

\title{Anomalous Entanglement in Chaotic Dirac Billiards}

\author{J. G. G. S. Ramos$^1$, I. M. L. da Silva$^2$, A. L. R. Barbosa$^2$ }

\affiliation{$^1$ Departamento de F\'isica, Universidade Federal da Para\'iba, 58051-970 Joa\~ao Pessoa  Para\'iba, Brazil\\
$^2$ Departamento de F\'{\i}sica,Universidade Federal Rural de Pernambuco, 52171-900 Recife, Pernambuco, Brazil}

\date{\today}

\begin{abstract}

We present analytical and numerical results that demonstrate the presence of anomalous entanglement behavior on the Dirac Billiards. We investigate the statistical distribution of the characteristic entangled measures, focusing on the mean, on the variance and on the quantum interference terms. We show a quite distinct behavior of the Dirac Billiard compared with the non-relativist (Schr\"odinger) ones. Particularly, we show a very plausible Bell state and a sharp amplitude of quantum interference term on entangled electrons left from the Dirac Billiards. The results have remarkable relevance to the novel quantum dots build of materials like graphene or topological insulators.

\end{abstract}
\maketitle

The entanglement is one the most fundamental effect on quantum mechanics, with no classical analog \cite{Sakurai}. Two or more particles are entangled if they support a nonlocal correlation which cannot be acquired by the dynamics of classical mechanics \cite{Alber}. Because its non-classical characteristics, the control of entangled states has attracted the interest of numerous science communities \cite{Horodecki}. The technological applications of the effect has a broad range as quantum computation, teleportation, telecommunication and cryptography \cite{Alber,Horodecki,Beenakker}.

A large number of mechanism to entangle electronic particles, with or without interaction, can be found in the literature\cite{Beenakker,Buttiker,Burkard}, and the quantum chaotic devices are a promising option \cite{BeenakkerEmary, AletaGopar}. In a recent work, Beenakker {\it et al.} \cite{BeenakkerMarcus} proposed the possibility to entangle two non-interacting electrons using, as an orbital entangler, a quantum (Schr{\"o}dinger) chaotic billiard or, as we will call henceforth, the Schr{\"o}dinger Billiard (SB). They obtained the averages and variances of concurrence and entanglement. However the results were found to be nearly invariant under the presence or absence of time-reversal symmetry (TRS). This means that the quantum interference corrections (weak-localization) of two arbitrary entanglement measures are approximately null, in contrast with another physical observables of electronic transport as conductance and shot-noise power. Nevertheless, the Ref. [\onlinecite{Gopar}] has argued that the two first moments of entanglement measures do not capture the full information about quantum dynamics, and the complete information about fundamental symmetries of nature emerges only on the distribution probabilities. More recently, the effects of tunneling barriers\cite{Souza,Vivo} on statistic of concurrence and joint probability distribution of concurrence and squared norm\cite{Vivo,Novaes} for SB were also studied.

The novel materials, including Dirac materials \cite{Wehling} (graphene and topological insulators), introduce new fundamental symmetries\cite{beenakkergrafeno,JacquodButtiker} that can affect the entangled electrons \cite{Kindermann}. In a recent investigation, the Ref.[\onlinecite{Barros}] analyzes how the sublattices or chiral symmetries \cite{Verbaarschot} (SLS) affect the electronic transport through a chaotic quantum (Dirac) billiard \cite{Barangergrafeno,Stampfer}, which henceforth we call chaotic Dirac Billiard (DB)\cite{Geim}. Through this device, the wave functions of the electrons are described by massless Dirac equation of the corresponding relativistic quantum mechanics, instead of Schr{\"o}dinger equation. The two categories of billiards show blunt differences on the corresponding electronic transport statistics, which leads us to suspect that a subtle difference occurs also on the quantum entanglement statistical moments.

\begin{figure}[!]
\includegraphics[width=7cm,height=9cm]{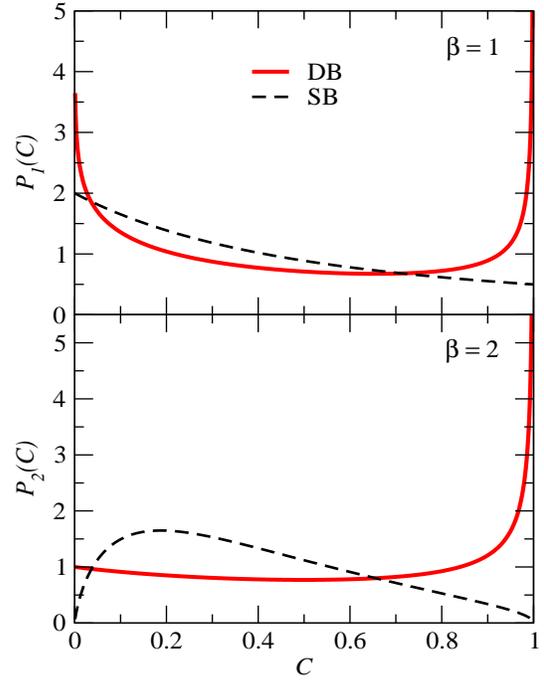} \quad \quad \quad
\caption{\scriptsize The distributions of concurrence ($\mathcal{P}_\beta(\mathcal{C})$) for the DB with ($\beta=1$) or without ($\beta=2$) TRS. The Eqs.(\ref{pc1}) and (\ref{pc2}) are depicted together with the corresponding ones for the SB \cite{Gopar}.}\label{figura1}
\end{figure}

In this work, we analyze the concurrence and entanglement statistics of two non-interacting electrons, with coherent phase transport, inside a chaotic DB. Two leads, both with two open channels, connects the DB with electronic source and drain at the Dirac point\cite{beenakkergrafeno,Barros}. We calculated the exacts distributions of concurrence ($\mathcal{C}$) for systems with or without TRS. We show a drastic dissimilarity on the characteristic distributions of the DB compared with the SB results \cite{Gopar}. We highlight the two main difference between DB and SB focusing on plausible experiments. Particularly, first we show that it is more probable the production of maximally entangled electronic state ($\mathcal{C}=1$ or Bell state) than of separable state ($\mathcal{C}=0$) for experiment using a DB, precisely the opposite to the behavior of SB, as indicated in Fig.(\ref{figura1}). Secondly, we show a weak localization correction of concurrence with the same order of magnitude of the mean (semiclassical) term, while for a SB it is approximately null\cite{BeenakkerMarcus}. Clearly, we indicate that the average and the variance of concurrence carry relevant information about quantum mechanics statistics of the DB, unlike what happens in SB. Our analytical results are in accordance with numerical simulation from the random matrix theory\cite{Verbaarschot,Barros, Macedo}.

{\it Scattering Model} - The setup consists of a DB connected to two leads, both with two open channels. We consider the system in the absence of electron-electron interactions and at zero temperature. The DB is represented by the massless Dirac Hamiltonian with SLS\cite{JacquodButtiker}. The Hamiltonian satisfy the following anti-commutation relation \cite{JacquodButtiker,Verbaarschot}
\begin{eqnarray}
\mathcal{H}= -\sigma_{z}\mathcal{H}\sigma_{z}, \quad
\sigma_{z}=
\left[
\begin{array}{cc}
\textbf{1}_{M} & 0\\
0 & -\textbf{1}_{M}
\end{array}
\right].\label{H}
\end{eqnarray}
The $\mathcal{H}$-matrix has dimension $2 M\times 2 M$, with $\textbf{1}_{M}$ denoting a $M \times M$ identity matrix. We can interpret the $M$ number of $1$'s and $-1$'s in $\sigma_z$ as the number of atoms in the sublattices $A$ and $B$ respectively\cite{beenakkergrafeno}, in a total of $2M$ atoms in the DB. The Hamiltonian model for the scattering matrix process can be written as\cite{Weidenmuller}
\begin{eqnarray}
\mathcal{S}(\epsilon)=\textbf{1}-2\pi i\mathcal{W}^{\dagger}(\epsilon-\mathcal{H} +i\pi\mathcal{W}\mathcal{W}^{\dagger})^{-1}\mathcal{W}.\label{MW}
\end{eqnarray}
The $\mathcal{S}$-matrix has dimension $4 \times 4$, indicating a total of two open channels in each terminal, each one originated from $A$ or $B$ sub-lattices. The $2M \times 4$ matrix $\mathcal{W}$ represents all deterministic couplings of the DB resonances to the open channels of the two terminals. The scattering matrix is unitary $\mathcal{S}^\dagger\mathcal{S}=\textbf{1}$ due to the conservation of electronic charge. It is convenient to represent the $\mathcal{S}$-matrix as a function of transmission, $t$, and reflection, $r$, blocks as
\begin{eqnarray}
\mathcal{S}=
\left[\begin{array}{cccc}
r&t'\\
t& r'
\end{array}\right],\label{Str}
\end{eqnarray}
where $t$, $t'$, $r$ and $r'$ have dimension $2\times 2$. From Eqs.(\ref{H}) and (\ref{MW}), the $\mathcal{S}$-matrix also satisfies the relation
\begin{eqnarray}
\mathcal{S}=\Sigma_{z}\mathcal{S}^{\dagger}\Sigma_{z}, \quad
\Sigma_{z}=
\left[\begin{array}{cc}
\textbf{1}_{2} & 0\\
0 & -\textbf{1}_{2}
\end{array}
\right] \label{S},
\end{eqnarray}
at the Dirac point, zero energy ($\epsilon =0$). Substituting the Eq.(\ref{S}) in Eq. (\ref{Str}), we conclude that $r=r^\dagger$, $r'=r'^\dagger$ and $t'=-t^\dagger$. 

We follow the previously mentioned RMT model for the ${\mathcal H}$ matrix and, consequently, for the ${\mathcal S}$ matrix to investigate the entanglement between two no interacting electrons as described in Refs.[\onlinecite{BeenakkerEmary,BeenakkerMarcus}]. Firstly, the concurrence of two electrons, after they left the mesoscopic device with SLS, can be written in terms of the transmission, $tt^{\dag}$, eigenvalues, $\tau_1$ and $\tau_2$, both encoded in the scattering matrix, Eqs.(\ref{Str}) and (\ref{S}), as
\begin{eqnarray}
\mathcal{C}=2\frac{\sqrt{\tau_1(1-\tau_1)\tau_2(1-\tau_2)}}{\tau_1+\tau_2-2\tau_1\tau_2}.\label{C}
\end{eqnarray}

The electronic states, after they left the DB, are separable or non entangled if $\mathcal{C}=0$, while, if $\mathcal{C}=1$, the particles are maximally entangled (Bell state). For intermediate values of $\mathcal{C}$ ​​between $0$ and $1$, the states are known as non-separable or partly entangled\cite{Souza}. The entanglement and concurrence are related through the equation\cite{Wootters}
\begin{eqnarray}
\varepsilon(\mathcal{C})=h\left(\frac{1+\sqrt{1-\mathcal{C}^2}}{2}\right)\label{E},
\end{eqnarray}
with
\begin{eqnarray}
h(x)=-x \log_2(x) - (1-x)\log_2(1-x).
\end{eqnarray}

{\it Probability Distributions} - To obtain the distribution of DB's concurrence, we start using the joint distribution, defined for the two characteristic eigenvalues, which was calculated in Ref.[\onlinecite{Macedo}] for $\beta=\{ 1,2\}$. The result can be written as
\begin{eqnarray}
\mathcal{P}_\beta(\phi_1,\phi_2)&=&c_{\beta} |\sin(\phi_1+\phi_2)\sin(\phi_1-\phi_2)|^{\beta} \nonumber\\&\times&\sin^{\beta-1}(2\phi_1)\sin^{\beta-1}(2\phi_2). \label{pphi}
\end{eqnarray}
The relation between the variables $\phi_i$ and the transmission eigenvalues $\tau_i$ is $\tau_i=\sin^2(2\phi_i)$, with $\phi_i \in [o,\pi/2]$, while normalization constants assume the values $c_1=1$ and $c_2=6$.

The Eqs.(\ref{C}) and (\ref{pphi}) can be used to calculate the concurrence distribution $\mathcal{P}_\beta(\mathcal{C})$ following the definition
\begin{eqnarray}
\mathcal{P}_\beta(\mathcal{C})&=&\left\langle\delta\left[\mathcal{C}-2\frac{\sqrt{\tau_1(1-\tau_1)\tau_2(1-\tau_2)}}{\tau_1+\tau_2-2\tau_1\tau_2}\right] \right\rangle,\label{pcdelta}
\end{eqnarray}
with $\left\langle\dots\right\rangle$ denoting the ensemble average performed with the eigenvalues distributions of Eq. (\ref{pphi}). The Eq. (\ref{pcdelta}) can be rewritten in the integral form as
\begin{eqnarray}
\mathcal{P}_\beta(\mathcal{C})&=&\int\int\delta\left[\mathcal{C}-2\frac{\tan(2\phi_1)\tan(2\phi_2)}{\tan^2(2\phi_1)+\tan^2(2\phi_2)}\right]\nonumber\\
&\times&\mathcal{P}_\beta(\phi_1,\phi_2)d\phi_1 d\phi_2 \label{pcphi}
\end{eqnarray}
\begin{figure*}[!]
\includegraphics[width=15cm,height=10cm]{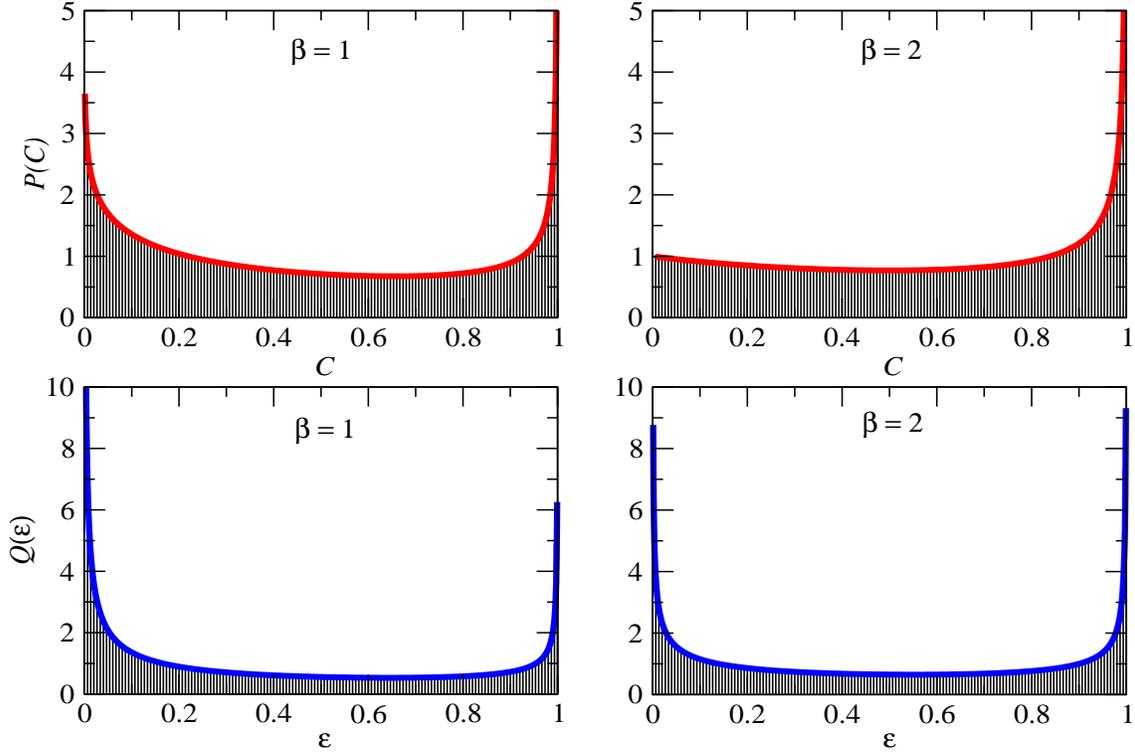} 
\caption{\scriptsize The distributions of concurrence ($\mathcal{P}_\beta(\mathcal{C})$) and entanglement ($\mathcal{Q}_\beta(\varepsilon)$) from Eqs.(\ref{pc1}), (\ref{pc2}) and (\ref{Q}) are depicted (continuous lines) for systems with ($\beta=1$) and without ($\beta=2$) TRS. The histogram is the numeric simulations of the Hamiltonian model with $10^6$ realizations of the corresponding random $\mathcal{S}$-matrix, the Eq. (\ref{MW}). The analytical and numerical results agrees nicely.}\label{grafico2}
\end{figure*}

The double integral over variables $\phi_1$ and $\phi_2$ in Eq. (\ref{pcphi}) can be performed through the transformation of variables $z_i=\tan^2(2\phi_i)$ and an appropriate expansion the delta function \cite{Vivo}
\begin{eqnarray}
\delta\left[\mathcal{C}-2\frac{\tan(2\phi_1)\tan(2\phi_2)}{\tan^2(2\phi_1)+\tan^2(2\phi_2)}\right]&=& \mathcal{F}_+(\mathcal{C}) z_2\delta\left[z_1-f_+(\mathcal{C}) z_2\right]\nonumber\\&+&\mathcal{F}_-(\mathcal{C}) z_2\delta\left[z_1-f_-(\mathcal{C}) z_2\right]\nonumber
\end{eqnarray}
with
\begin{eqnarray}
\mathcal{F}_\pm(\mathcal{C})&\equiv&\frac{\left(f_\pm(\mathcal{C})+1\right)^2\sqrt{f_\pm(\mathcal{C})}}{\pm f_\pm(\mathcal{C})\mp 1};
\nonumber\\
f_\pm(\mathcal{C})&\equiv &\frac{2-\mathcal{C}^2\pm 2 \sqrt{1-\mathcal{C}^2}}{\mathcal{C}^2}.
\nonumber
\end{eqnarray}

After a some algebra, we obtain the following expressions for the probability distribution of concurrence with preserved TRI ($\beta=1$)
\begin{eqnarray} \label{pc1}
\mathcal{P}_1(\mathcal{C})&=&\frac{1}{2}\frac{1}{1-\mathcal{C}^2}\left\{\frac{\sqrt{2}}{2}\sqrt{1+\frac{\sqrt{1-\mathcal{C}^2}}{1-\mathcal{C}^2}}\right. \\
&\times&\left.\textrm{arccoth}\left[ {\sqrt{1+\frac{\mathcal{C}^2}{2\left(1-\mathcal{C}^2\right)\left(1+\frac{\sqrt{1-\mathcal{C}^2}}{1-\mathcal{C}^2}\right)}}}\right]-1 \right\}\nonumber
\end{eqnarray}
and, for broken TRI ($\beta=2$), it renders a simple expression
\begin{eqnarray}
\mathcal{P}_2(\mathcal{C})&=&\frac{\sqrt{1-\mathcal{C}^2}}{(1+\mathcal{C})(1-\mathcal{C}^2)}. \label{pc2}
\end{eqnarray}

The Fig.(\ref{figura1}) shows the distributions of concurrence, Eqs. (\ref{pc1}) and (\ref{pc2}), for the DB. We plot in the same place the distributions for the SB, which were obtained in Ref.[\onlinecite{Gopar}], for a direct comparison. The systems exhibit a quite distinct behavior. We observe, as the main characteristic of DB, a major probability to find the electrons in the maximally entangled states ($\mathcal{C}=1$), contrary to the behavior of the SB. However, the preserved TRS supports electrons with the probability of separable states ($\mathcal{C}=0$) significantly increased, as depicted in the Fig.(\ref{figura1}).
The mean value of concurrence can be obtained from Eqs.(\ref{pc1}) and (\ref{pc2}) and renders
\begin{eqnarray}
\left\langle\mathcal{C}\right\rangle\approx\left\{ \matrix{
0.4675 \qquad \beta = 1 \cr
0.5708 \qquad \beta = 2 }\right.
\label{mean}
\end{eqnarray}
Using Eq.(\ref{mean}), the weak localization of concurrence can be calculated with the same algebra and renders
\begin{eqnarray}
\left\langle\mathcal{C}\right\rangle_{wl}=\left\langle\mathcal{C}\right\rangle_{1}-\left\langle\mathcal{C}\right\rangle_2\approx-0.1033.
\label{wl}
\end{eqnarray}
Notice the weak localization term has the same magnitude order of the main (semiclassical) term, Eq.(\ref{mean}), while for the SB it is approximately null\cite{BeenakkerMarcus}. The TRI, unlike what happens in SB, has a strong influence over average of concurrence in DB. 
For the variance of concurrence we obtain
\begin{eqnarray}
\textrm{var}\left[\mathcal{C}\right]\approx\left\{ \matrix{
0.1117 \qquad \beta = 1 \cr
0.1034 \qquad \beta = 2 }\right.
\label{var}
\end{eqnarray}
Like weak localization, the variance of concurrence has the same order of magnitude of the its mean, Eq.(\ref{mean}).
Lastly, we obtain the distribution of entanglement $\mathcal{Q}_\beta$ from Eqs. (\ref{pc1}) and (\ref{pc2}) using the following appropriate change of variables \cite{Gopar}
\begin{eqnarray}
\mathcal{Q}_\beta(\varepsilon)&=&\frac{\ln (2)}{\mathcal{C}(\varepsilon)}\frac{\sqrt{1-\mathcal{C}(\varepsilon)^2}}{\textrm{arctanh} [\sqrt{1-\mathcal{C}(\varepsilon)^2}]}\mathcal{P}_\beta(\varepsilon). \label{Q}
\end{eqnarray}
The analytical results for the distributions of entanglement are cumbersome, but it is depicted in the Fig.(\ref{grafico2}). The average of entanglement renders the results
\begin{eqnarray}
\left\langle\varepsilon\right\rangle\approx\left\{ \matrix{
0.382 \qquad \beta = 1 \cr
0.485 \qquad \beta = 2 }\right.
\label{meane}
\end{eqnarray}
while its weak localization amplitude term is
\begin{eqnarray}
\left\langle\varepsilon\right\rangle_{wl}=\left\langle\varepsilon\right\rangle_{1}-\left\langle\varepsilon\right\rangle_2\approx-0.10.
\label{wle}
\end{eqnarray}
For the variance of entanglement we obtain
\begin{eqnarray}
\textrm{var}\left[\varepsilon\right]\approx\left\{ \matrix{
0.121 \qquad \beta = 1 \cr
0.122 \qquad \beta = 2 }\right.
\label{vare}
\end{eqnarray}

{\it Numeric Simulation} - In order to confirm the analytical results, the Eqs.(\ref{pc1}) and (\ref{pc2}), we employ a numerical simulation using the $\mathcal{S}$-matrix formulation of Eq.(\ref{MW}). The ensemble of Hamiltoninans satisfies the SLS of Eq.(\ref{H}). The numerical simulation was developed in Ref.[\onlinecite{Barros}]. The anti-commutation relation, Eq.(\ref{H}), implies a Hamiltonian member of the ensemble rewritten as\cite{Verbaarschot}
\begin{eqnarray}
\mathcal{H}=\left(
\begin{array}{cc}
\textbf{0} & \mathcal{T} \\
\mathcal{T}^{\dagger} & \textbf{0}
\end{array}
\right).
\label{H1}
\end{eqnarray}
Here, the $ \mathcal{T} $-matrix block of the $\mathcal{H}$- matrix has dimension $M\times M$. The Random Matrix Theory establishes that the entries of $\mathcal{T}$-matrix can be chosen as a member of a Gaussian distribution \cite{Macedo}
\begin{eqnarray}
P( \mathcal{T} )\propto \exp\left\lbrace -\frac{\beta M}{2\lambda^{2}}Tr( \mathcal{T} ^{\dagger} \mathcal{T} )\right\rbrace,
\end{eqnarray}
where $2M$ is the number of resonances inside the DB, including both the two sub-lattices degrees of freedom. Also, $\lambda=2M\Delta/\pi$ is the variance, related to the electronic single-particle level spacing, $\Delta$. The $\mathcal{W} = \textbf{(}\mathcal{W}_{1}, \mathcal{W}_{2}\textbf{)} $ matrix is a $2M \times 4$ deterministic matrix, describing the coupling of the resonances states of the chaotic DB with the propagating modes in the two terminals. This deterministic matrix satisfies non-direct process, i.e., the orthogonality condition $ \mathcal{W}_{p}^{\dagger}\mathcal{W}_{q}=\frac{1}{\pi}\delta_{p,q}$ holds. Accordingly, we consider the relation $\sigma_z \mathcal{W}\Sigma_z=\mathcal{W}$, indicating the scattering matrix is symmetric (\ref{S}). We consider the system on the Dirac point, $\epsilon=0$, and, to ensure the chaotic regime and consequently the universality of observables, the number of resonances inside DB is large ($M\gg 4$). 

The numerical simulations produce the Fig.(\ref{grafico2}), which shows the distributions of concurrence and entanglement obtained
through $10^6$ realizations compared with the analytical results,
Eqs.(\ref{pc1}), (\ref{pc2}) and (\ref{Q}), for systems with or without TRS. We use the $\mathcal{T} $ matrices, with dimension $100\times 100$ ($M=100$), and the corresponding $\mathcal{H}$ matrices with dimension $200 \times 200$ ($200$ resonances).

{\it Conclusions} - We present a complete statistical study of the entangled electronic measures. We show a very peculiar quantum behavior if the production of orbital entanglement is raised on the Dirac billiards quantum dots. The analytical expressions for the distributions of concurrence were obtained in the presence or absence of TRS. We compare the results for the Dirac billiards with previous ones of the Schr\"odinger billiards, Refs. [\onlinecite{BeenakkerMarcus,Gopar}], and found significant statistics differences between these devices. Clearly, we indicate that the average, the variance of concurrence and entanglement carry relevant information about quantum mechanics statistics of the DB, unlike what happens in SB. Finally, our analytical results are confirmed by numeric simulation from random matrix theory.

This work was partially supported by CNPq, CAPES and FACEPE (Brazilian
Agencies).

\end{document}